# Search for cosmic gamma rays with the Carpet-2 extensive air shower array

D.D. Dzhappuev, V.B. Petkov, A.U. Kudzhaev, N.F. Klimenko, A.S. Lidvansky, S.V. Troitsky

*Institute for Nuclear Research, Russian Academy of Sciences; Moscow, Russia;*

*dzhappuev@mail.ru*

**Abstract**   The present-day status of the problem of searching for primary cosmic gamma rays at energies above 100 TeV is discussed, as well as a proposal for a new experiment in this field. It is shown that an increase of the area of the muon detector of the Carpet-2 air shower array up to 410 square meters, to be realized in 2016, will make this array quite competitive with past and existing experiments, especially at modest energies. Some preliminary results of measurements made with smaller area of the muon detector are presented together with estimates of expected results to be obtained with the coming large-area muon detector.

## 1. Introduction

Search for primary gamma rays of energies higher than 100 TeV using the extensive air shower (EAS) method started in 1960s. A lot of experiments were made in this line of research until the present time, different types of detectors and different methods of isolating showers produced by primary gamma rays being used. One can apply the EAS detection method to studies of the diffuse gamma ray emission of cosmic origin, if there is a way of efficient separation of showers produced by primary photons from EAS generated by primary protons and nuclei. Such a separation is possible due to the fact that the showers from primary photons are substantially less abundant with hadrons (and, consequently, muons) than proton showers (the more so in case of showers from nuclei). Thus, by selecting EAS with diminished number of hadrons or muons one can hope to isolate the showers from primary gamma rays. Maze and Zawadski [1] were the first who put forward the idea of searching for high-energy gamma rays by way of detecting muon-poor showers. This seemed to be the simplest way of distinguishing gamma-ray induced showers from ordinary extensive air showers. Following this idea, several groups tried to measure the flux of diffuse gamma ray emission and claimed to obtain some positive results. The experiments at Mt. Chacaltaya [2] and Tien Shan [3], in Yakutsk [4] and Lodz [5] published such results, but they have insufficient statistical significance and were not confirmed later. More careful subsequent experiments (collaborations EAS-TOP [6], CASA-MIA [7] and KASKADE [8] in the energy range $3\cdot10^{14} - 5\cdot10^{16}$ eV, and Haverah Park [9], AGASA [10]−[12], Yakutsk [13], [14], Pierre Auger [15], [16], and Telescope Array [17] at energies higher than $10^{18}$ eV) yielded only upper limits on the fluxes of cosmic gamma rays.

## 2. Present-day Status of the Problem

The upper limits obtained in [6]–[17] are significantly lower than the fluxes measured in early works [2]–[5] (see Fig. 1). Quite recently, new interesting results have appeared in the energy range $5 \cdot 10^{15} - 2 \cdot 10^{17}$ eV. Archival data of the MSU air shower array were analyzed in [18], [19]. Showers in this experiment were selected according to the low content of muons with energies higher than 10 GeV. For the entire energy range under analysis only upper limits on the cosmic gamma ray flux were obtained, excluding a narrow interval $5 \cdot 10^{16} - 10^{17}$ eV, where muonless showers were recorded, whose number well exceeded the expected number of background events. This allowed authors to estimate the absolute value of the flux of diffuse gamma rays in this energy range. These results are in line with the most recent KASCADE-Grande flux limits at these energies [20].

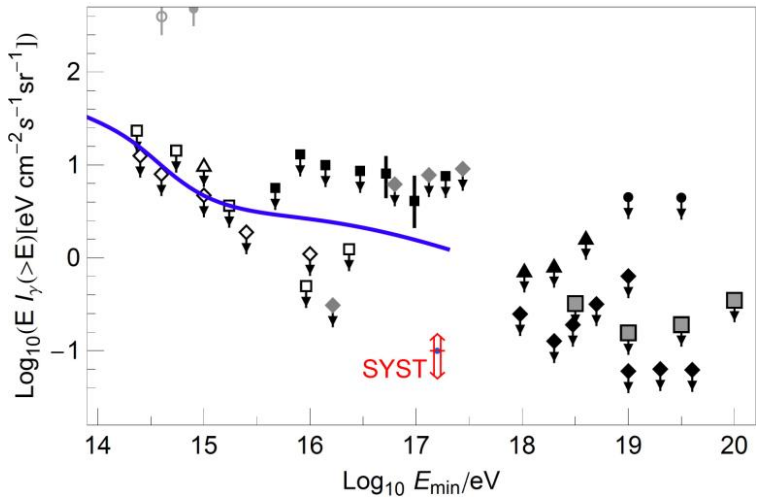

**Fig. 1.** *Estimates of the integral gamma-ray flux: Detection claims by Tien Shan (gray open circle) and Lodz (gray solid circle), and EAS-MSU (black squares and error bars). Open triangles, squares and diamonds are for EAS-TOP, CASA-MIA, and KASCADE, respectively; gray diamonds: KASCADE-Grande, black triangles: Yakutsk, black diamonds: Pierre Auger, small black circles: AGASA, large squares: Telescope Array. The curve represents an example theoretical prediction [23] for the model in which photons and neutrinos are produced in cosmic-ray collisions with the hot gas surrounding our Galaxy, assuming the best-fit IceCube observed neutrino spectrum.*

It should be noted that a new impetus to interest in this problem was given by publications of the IceCube results on detection of high-energy astrophysical neutrinos [21], [22]. Neutrinos produced in decays of charged pions should be accompanied by gamma rays produced in decays of neutral pions. Hence, there is a motivation for new specialized experiments with more precise measurements of the flux of diffuse gamma rays at energies higher than 100 TeV.

The Carpet-2 air shower array of the Baksan Neutrino Observatory includes a large area muon detector that in principle is capable of separating the showers from primary gamma rays with energies higher than 100 TeV. In this paper we present preliminary estimates of sensitivity of the array to the diffuse flux of cosmic gamma rays with the existing muon detector and analyze possible results that can be achieved after a good exposure with the increased area of the muon detector after its modernization (this work is now in progress).

## 3. Experiment

The Carpet-2 air shower array [24], [25] consists of a surface part (the original Carpet with six external huts) and underground muon detector (MD). The distance between centers of the Carpet and MD is 47 m. The Carpet that detects the EAS electron-photon component includes 400 scintillation detectors forming a square (20x20) with a total area of 196 m$^2$. The muon detector records the muon component with energy threshold of 1 GeV. Signals from external huts, each of which contains 9 m$^2$ of scintillation detectors, are used for determination of shower arrival directions. The accuracy of determination of coordinates of EAS axes hitting the Carpet is no worse than 0.7 m, while the arrival directions of showers are measured with an accuracy of better than ~ 3°. The Carpet and MD operate independently of each other and have different dead times of recording electronics. But time markers of events in the MD and Carpet are produced by one and the same clock, so that coincident events are identified within a time interval $\Delta t = 1$ ms. The total number of relativistic particles within the Carpet ($N_{r.p.}$) and the number $n_\mu$ of muons recorded by the MD are the experimentally measured quantities used to determine the energy of EAS and the total number of muons in it, respectively. The events satisfying the following conditions are included into processing:
1. shower axes are well within the Carpet;
2. zenith angles of showers $\theta < 40°$;
3. $N_{r.p.} \geq 2 \cdot 10^4$.

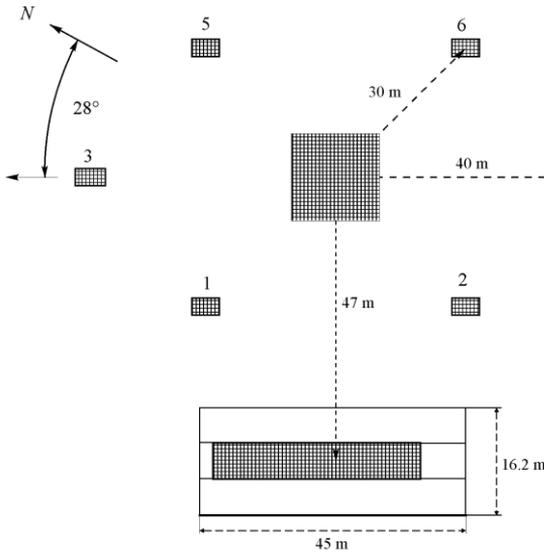

**Fig. 2.** *The layout of the Carpet-2 air shower array.*

### 3.1. Simulations

For the sensitivity estimates presented here, the CORSIKA code v. 6720 (QGSJET01C, FLUKA 2006) [26] was used for modeling the showers. 5400 showers from primary protons and 6597 showers from primary iron nuclei were simulated within the energy interval (0.316−31.6) PeV, as well as 815 showers from primary gamma rays in the range (0.3−9) PeV.

As a result of modeling the following averaged relations were obtained:

$E_p$ [GeV] = 174·$N_{r.p.}^{0.75}$ (1)

$E_\gamma$ [GeV] = 1096·$N_{r.p.}^{0.62}$ (2)

### 3.2. Preliminary analysis and future prospects

Figure 3 presents the calculated correlation distribution $n_\mu$ versus $N_{r.p.}$ for showers from primary protons, iron nuclei and gamma rays, where points with $n_\mu = 0$ are shown as $n_\mu = 0.2$. Though the numbers of simulated proton and gamma-ray showers are not equal, all muon-poor showers with $N_{r.p.} \geq 10^5$ are produced by gamma rays.

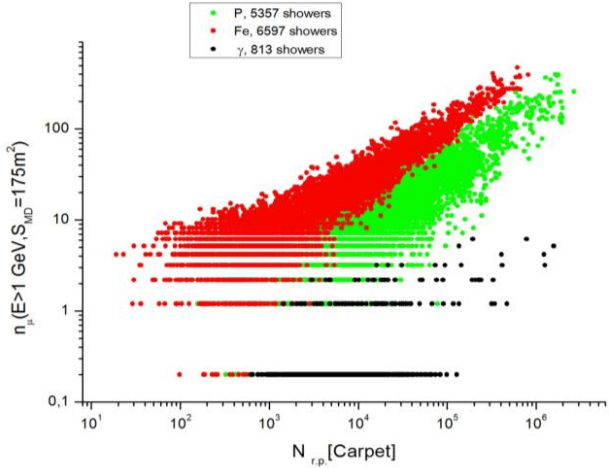

***Fig 3.*** *The calculated $n_\mu$ - $N_{r.p.}$ dependence (CORSIKA) for EAS induced by protons, iron nuclei and gamma rays.*

A similar distribution for experimental data is presented in Fig. 4 together with expectation for gamma rays. To compare the simulations with the data, a subset of 4261 showers recorded during 226 days of work of 175 m² MD was analyzed (the exposure was 5.74·10¹³ cm² s sr). Experimental muonless events (their number is 1080) are not shown in the figure. Of course, such a number of muonless showers is not necessarily produced by gamma rays, but can be a consequence of fluctuations of the number of muons or of incorrect method of selecting the events. Indeed, as follows from these two distributions the number of muons fluctuates strongly in the interval $N_{r.p.} = 2·10^4 \div 10^5$, so that it is impossible to isolate real gamma rays against the background proton showers at currently available statistics of experimental and simulated showers. Only the upper limit on the flux of primary gamma rays can be evaluated, assuming that the detected muonless showers are the background ones. Preliminary constraints on the flux of diffuse gamma rays derived by us are considerably worse as compared to the results obtained earlier in different experiments. In order to improve them we plan to process archival data accumulated for 10 years.

One can diminish relative fluctuations of the number of muons recorded by the MD (thus improving the efficiency of separation of showers) by increasing the muon detector area. This work is now in progress. Next year it is planned to install 235 new scintillation counters with area of 1 m² each in tunnels of the MD, thus increasing the MD area up to 410 m². According to further plans, the effective area of the array will be increased up to 6·10³ m² (the Carpet-3 array). The area of the MD will be increased in this case up to the maximum possible value of 615 m² (total filling of the three MD tunnels).

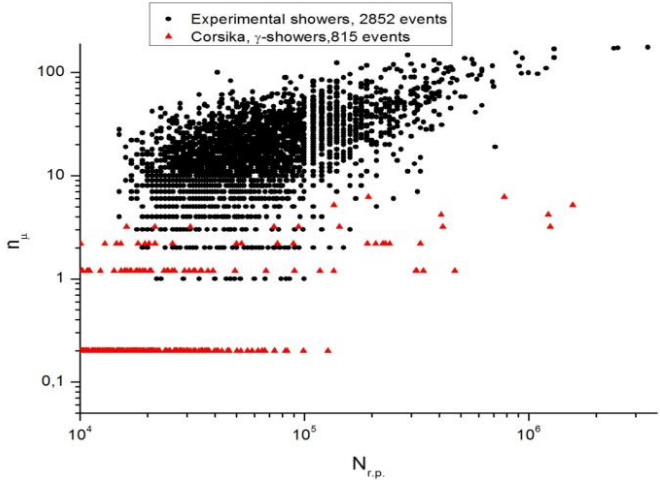

**Fig. 4.** *Measured and calculated nμ - Nr.p. dependence (experiment, CORSIKA)*

Commissioning of the 410 m² MD will become a crucial step towards gamma-ray astronomy with Carpet-2. Preliminary estimates demonstrate that the background of muuonless events from hadronic showers will be reduced drastically in this configuration. Figure 5 demonstrates the estimated sensitivity of the existing array (Carpet-2) and two its modernizations (Carpet-2+ and Carpet-3) as compared to the upper limits obtained by other experiments. One can see that using a new configuration of the Baksan array a significant advance can be achieved at energies below and around $10^{15}$ eV.

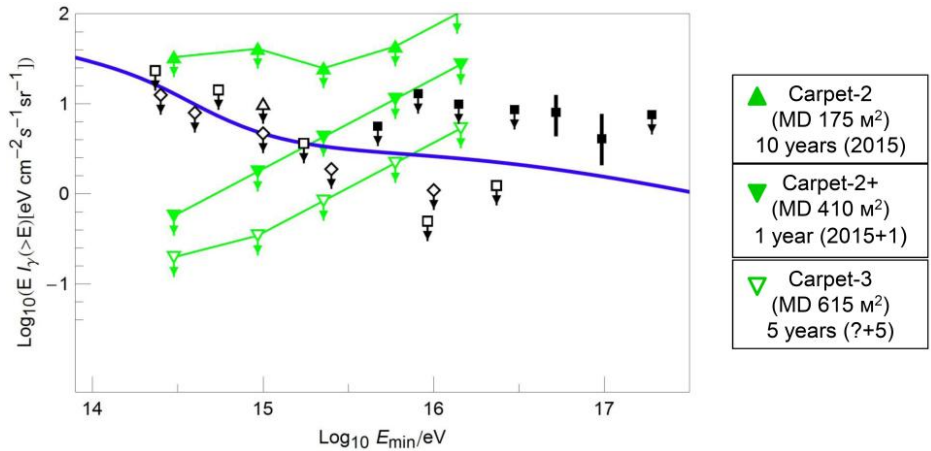

**Fig. 5.** *The sensitivity of the Carpet-2 and Carpet-3 air shower arrays to the diffuse cosmic photons.*

The sensitivity of the upgraded array to point sources will depend on the source's position in the sky. Preliminary estimates give the sensitivity of ~5·$10^{-13}$ cm⁻² s⁻¹ for the integral gamma-ray flux of Mrk 501 above 100 TeV (one year of observations with 410 m² MD). In more detail, the estimates of sensitivity of Carpet-2+ for gamma rays will be discussed in a forthcoming paper.

## 4. Conclusions

1. The Carpet air shower array could be quite competitive in gamma-ray astronomy above 100 TeV.

2. To perform this task better, it is highly desirable to increase the area of the Muon Detector as much as possible.

3. The work to reach an area of 410 $m^2$ next year and 615 $m^2$ in the future is now in progress.

## Acknowledgments

The work of DD, VP, AK, and AL is supported by the Russian Foundation for Basic Research, project number 16-02-0687. The work of ST related to testing the EAS-MSU result with different experimental data is supported by the Russian Science Foundation, grant no. 14-12-01340.